\documentclass{cimento}

\usepackage{graphicx}  

\title{Ultra High Energy Cosmic Rays, Photons and Neutrinos}
\author{Roberto Aloisio}
\instlist{\inst{ }Gran Sasso Science Institute, viale F. Crispi 7, I-67100 L'Aquila, Italy
\inst{ }INFN - Laboratori Nazionali Gran Sasso, via G. Acitelli 22, I-67010 Assergi (AQ), Italy.}

\begin{document}

\maketitle

\begin{abstract}
We briefly review the main channels of gamma ray and neutrino production due to the propagation of ultra high energy cosmic rays. We show how the observations of these secondary radiations are of paramount importance in tagging mass composition and sources of ultra high energy cosmic rays.
\end{abstract}

\section{Introduction}

Ultra High Energy Cosmic Rays (UHECR) are the most energetic particles observed in nature with energies up to $10^{20}$ eV. The most advanced experiments to detect UHECR are the Pierre Auger Observatory in Argentina \cite{ThePierreAuger:2015rma}, far the largest experimental setup devoted to the study of UHECR, and the Telescope Array (TA) experiment \cite{Tinyakov:2014lla}, placed in the United States. The experimental study of UHECR clarified few important characteristics of these particles: (i) UHECR are charged particles with a limit on photon and neutrino fluxes around $10^{19}$ eV at the level of few percent and well below respectively, (ii) the spectra observed at the Earth show a slight flattening at energies around $5\times 10^{18}$ eV (called the ankle) with (iii) a steep suppression at the highest energies.

The composition of UHECR is still matter of debate. Before the advent of Auger the experimental evidences were all pointing toward a light composition with a proton dominated flux until the highest energies, sources injecting soft spectra and acceleration energies larger than $10^{20}$ eV, the so-called dip model \cite{Aloisio:2006wv}. The measurements carried out by the Auger observatory \cite{Aab:2014kda} have shown that the mass composition of CRs, from prevalently light at $\sim 10^{18}$ eV, becomes increasingly heavier towards higher energies. Several independent calculations (see \cite{Aloisio:2013hya,Aab:2016zth,Unger:2015laa,Globus:2014fka} and references therein) showed that spectrum and composition observed by Auger can be well explained only if sources  provide heavy nuclei with very hard spectra and a maximum energy $\sim 5\times 10^{18}$ eV. If confirmed, these mixed composition models would represent a change of paradigm respect to the picture of ten years ago. On the other hand, the TA experiment, even if with $1/10$ of the Auger statistics, collected data that seem to confirm the pre-Auger scenario \cite{Abbasi:2014sfa}, the mass composition is compatible with being light for energies above $10^{18}$ eV, with no apparent transition to a heavier mass composition. 

A joint working group made of members of both collaborations, TA and Auger, has recently concluded that the results of the two experiments are not in conflict once systematic and statistical uncertainties have been taken into account \cite{Abbasi:2015xga}. This conclusion, though encouraging on one hand, casts serious doubts on the possibility of reliably measuring the mass composition at the highest energies, unless some substantially new piece of information becomes available. For this reason we consider here both alternatives of a pure proton composition (dip model) and mixed composition models. 

The extra-galactic origin of UHECRs, at least at energies above the ankle $E>10^{19}$ eV, is widely accepted \cite{Aloisio:2012ba}. The propagation of UHECR through the intergalactic medium is conditioned primarily by astrophysical photon backgrounds. The astrophysical backgrounds involved are the Cosmic Microwave Background (CMB) and the Extra-galactic Background Light (EBL). The interactions of UHECR (protons\footnote{Here we do not consider the case of neutrons because their decay time is much shorter than all other scales involved in the propagation of UHECR \cite{Aloisio:2008pp,Aloisio:2010he}.} or heavier nuclei) with astrophysical backgrounds give rise to the processes of: pair-production, photo-pion production and, only in the case of nuclei heavier than protons, photo-disintegration. Moreover, protons propagation is affected only by the CMB while for nuclei, and only in the case of photo-disintegration, also the EBL plays a role \cite{Aloisio:2008pp,Aloisio:2010he}.
	
These interactions shape the spectrum of UHECR observed at the Earth and are also responsible for the production of secondary (cosmogenic) particles: photons and neutrinos. This secondary radiation can be observed through ground-based or satellite experiments and brings important informations about the mass composition of UHECR and, possibly, on their sources. In the present paper we will briefly review the main theoretical expectations concerning cosmogenic neutrinos and gamma rays and their impact on our understanding of UHECR\footnote{The present work is the proceedings paper of a review talk, on cosmic rays and their connections with gamma rays and high energy neutrinos, given at the 11$^{{\rm th}}$ workshop on "Science with the New generation of High Energy Gamma-ray Experiments". Due to the lack of space, we will not touch here all the topics discussed in the oral presentation, focusing only on the production of secondary gamma-rays and neutrinos by the propagation of UHECR.}. 

\section{Neutrinos}

\begin{figure}[!h]
\centering
\includegraphics[scale=.36]{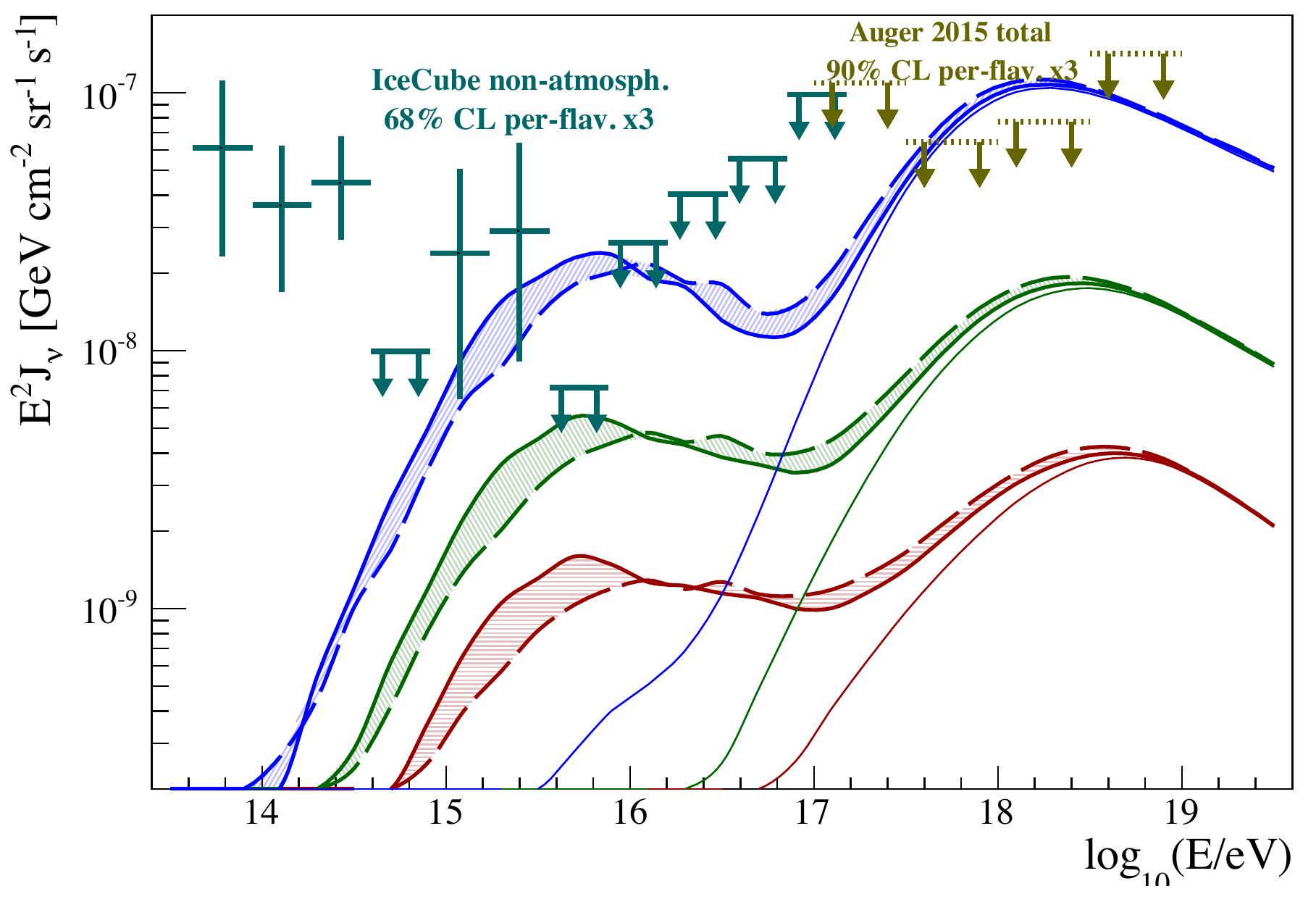}
\includegraphics[scale=.36]{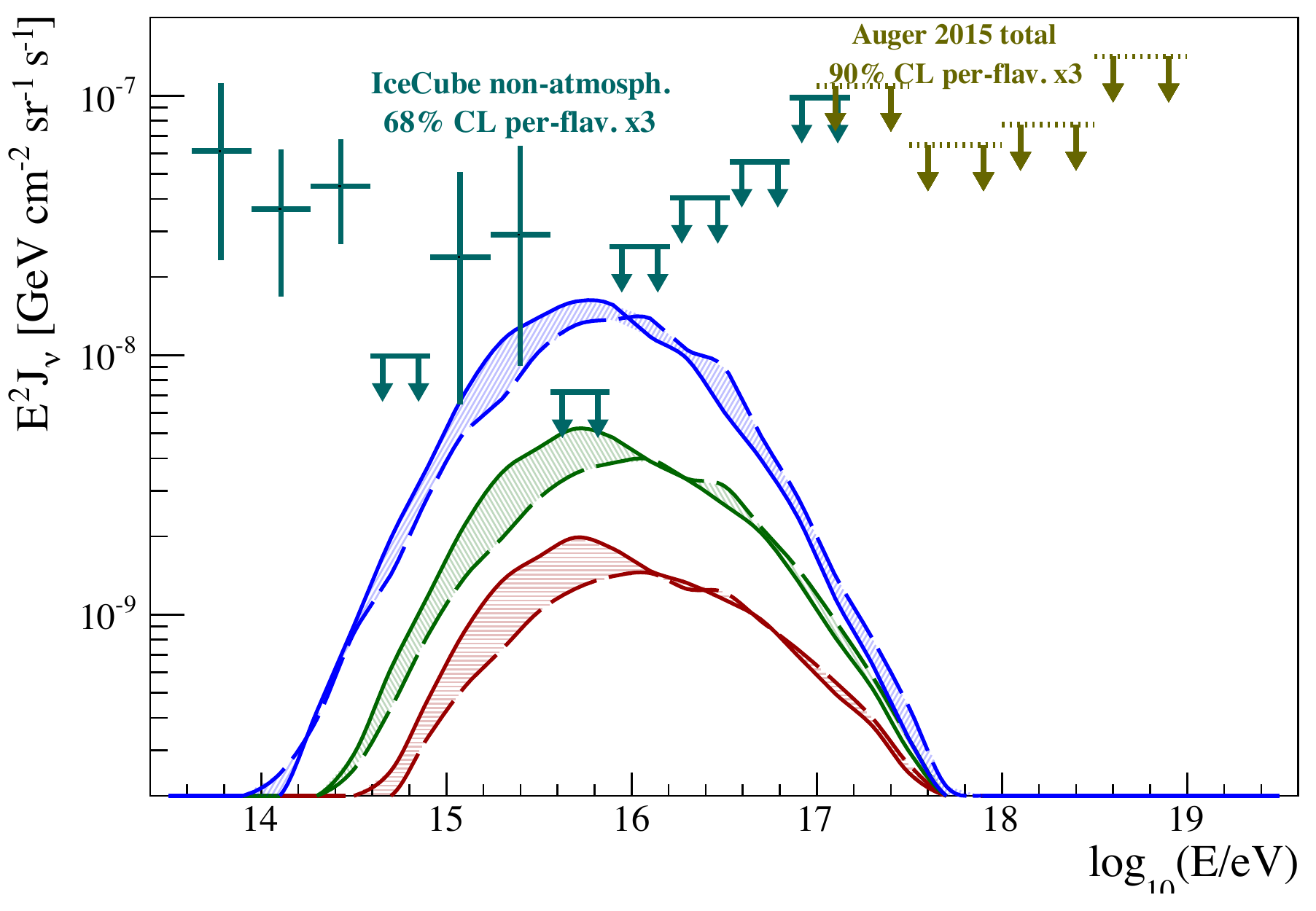}
\caption{[Left Panel] Fluxes of neutrinos in the case of the dip model. The three different fluxes correspond to different assumptions on the cosmological evolution of sources, coloured bands show the uncertainties due to the EBL model \cite{Aloisio:2015ega}. Thin solid lines are neutrino fluxes obtained taking into account the sole CMB field. [Right Panel] Neutrino fluxes in the case of mixed composition with the same color code of left panel. Experimental points are the observations of IceCube on extra-terrestrial neutrinos and the Auger limits on neutrino fluxes.} 
\label{fig8}   
\end{figure}

There are two processes by which neutrinos can be emitted in the propagation of UHECR: (i) the decay of charged pions, produced by photo-pion production, $\pi^{\pm}\to \mu^{\pm}+\nu_{\mu}(\bar{\nu}_{\mu})$ and the subsequent muon decay $\mu^{\pm}\to e^{\pm}+\bar{\nu}_{\mu}(\nu_{\mu})+\nu_e(\bar{\nu}_e)$; (ii) the beta-decay of neutrons and nuclei produced by photo-disintegration: $n\to p+e^{-}+\bar{\nu}_e$, $(A,Z)\to (A,Z-1)+e^{+}+\nu_e$, or $(A,Z)\to (A,Z+1)+e^{-}+\bar{\nu}_e$. These processes produce neutrinos in different energy ranges: in the former the energy of each neutrino is around a few percent of that of the parent nucleon, whereas in the latter it is less than one part per thousand (in the case of neutron decay, larger for certain unstable nuclei). This means that in the interaction with CMB photons, which has a threshold Lorentz factor of about $\Gamma\ge 10^{10}$, neutrinos are produced with energies of the order of $10^{18}$~eV and $10^{16}$~eV respectively. Interactions with EBL photons contribute, with a lower probability than CMB photons, to the production of neutrinos with energies of the order of $10^{15}$~eV in the case of photo-pion production and $10^{14}$~eV in the case of neutron decay (see \cite{Aloisio:2015ega} and references therein). The flux of secondary neutrinos is very much sensitive to the composition of UHECR. In figure \ref{fig8} (taken from \cite{Aloisio:2015ega}) we plot the flux of cosmogenic neutrinos expected in the case of the dip model (left panel) and in the case of mixed composition (right panel). Comparing the two panels of figure \ref{fig8} it is evident the huge impact of the composition on the expected neutrino flux: heavy nuclei provide a reduced flux of neutrinos because the photo-pion production process in this case is subdominant. 

The production of cosmogenic neutrinos is almost independent of the variations in sources' distribution because the overall universe, up to the maximum red-shift (of sources) $z_{max}\simeq 10$ \cite{Berezinsky:2011bb}, could contribute to the flux. Once produced at these cosmological distances neutrinos travel toward the observer almost freely (see \cite{Gondolo:1991rn} and references therein), except for the adiabatic energy losses and flavour oscillations. This is an important point that makes neutrinos a viable probe not only of the mass composition of UHECRs but also of the cosmological evolution of sources. In figure \ref{fig8} three different hypothesis on the cosmological evolution of sources are taken into account: no cosmological evolution (red bands), evolution typical of the star formation rate (green band) and of active galactic nuclei (blue band), see \cite{Aloisio:2015ega} and references therein. 
  
\begin{figure}[!h]
\centering
\includegraphics[scale=.35]{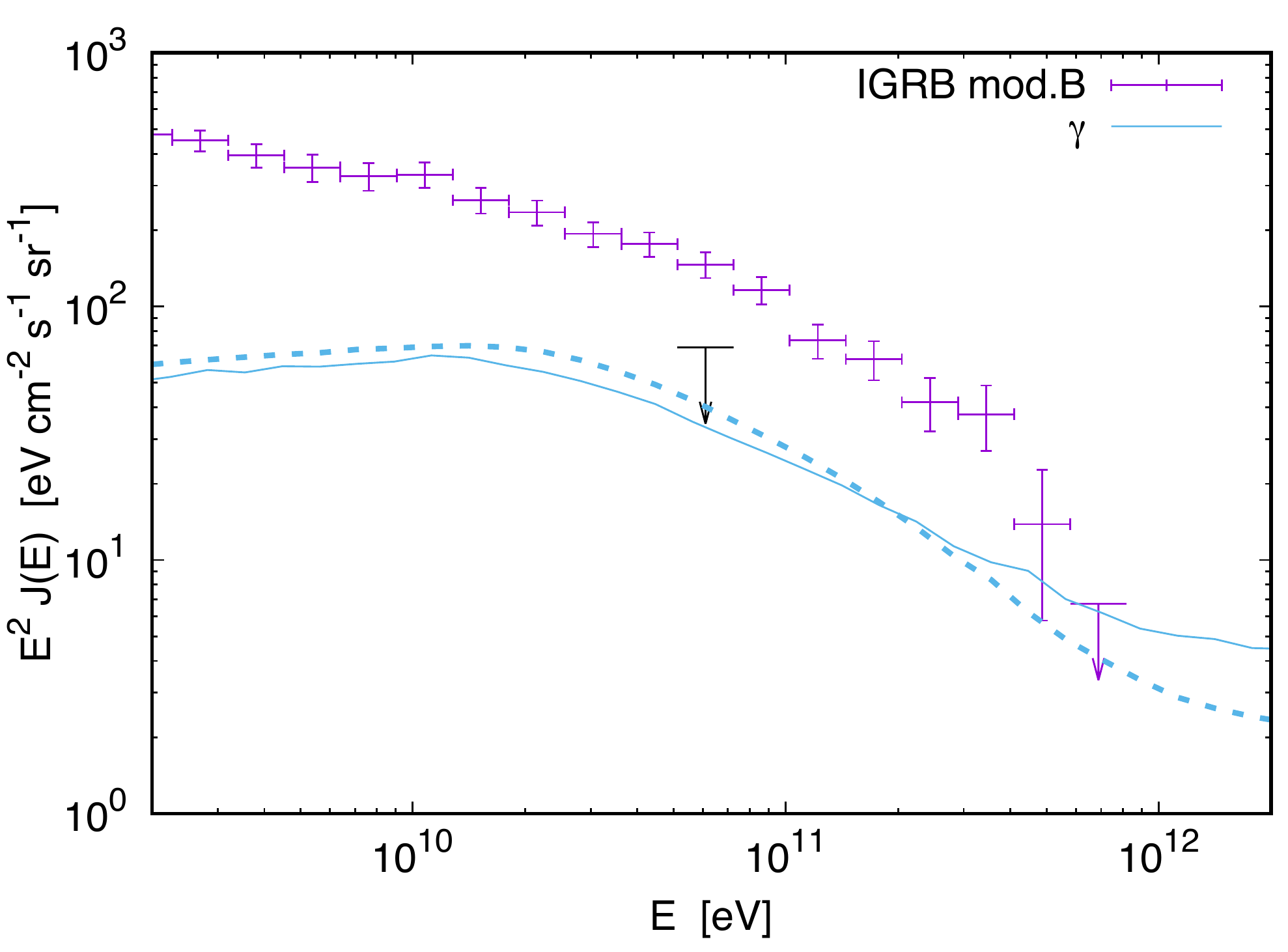}
\includegraphics[scale=.37]{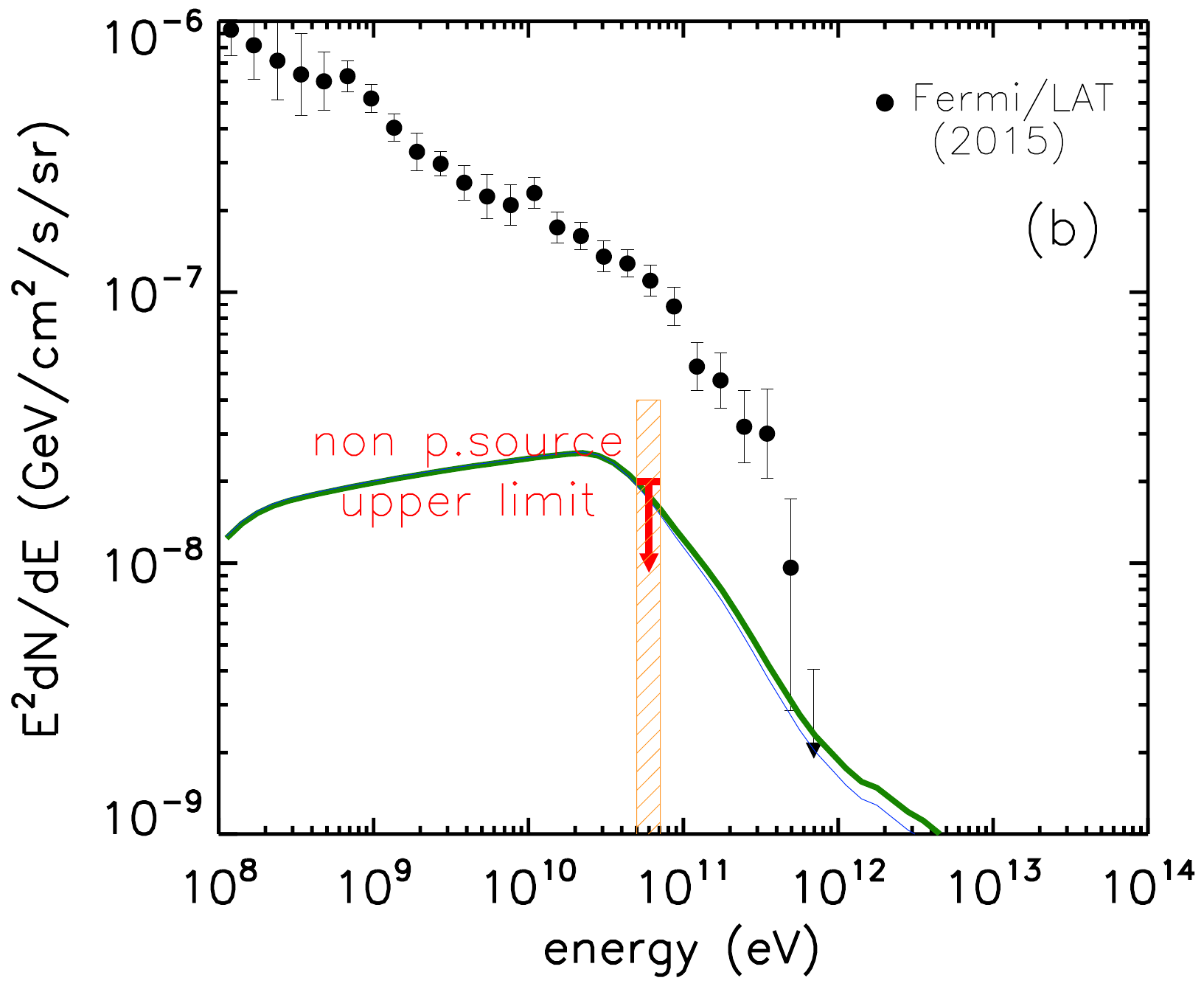}
\caption{Spectra of cosmogenic gamma rays obtained in the case of pure proton composition of UHECR without cosmological evolution of sources, as computed in \cite{Berezinsky:2016jys} (left panel) and in \cite{Liu:2016brs} (right panel), together with the Fermi-LAT data on diffuse gamma ray background, as in model-B (left panel) and model-A (right panel) of the analysis presented in \cite{TheFermi-LAT:2015ykq}.} 
\label{fig8bis}   
\end{figure}
    
There is a solid consensus about the light composition of UHECRs in the low energy part of the observed spectrum. This assures a flux of cosmogenic neutrinos in the PeV energy region, produced by protons photo-pion production on the EBL photons. Coloured bands in figure \ref{fig8} show the uncertainties connected with the EBL background \cite{Aloisio:2015ega}. Another important uncertainty in the expected neutrino flux comes from the contribution of UHECR sources at high red-shift. Given the energy losses suffered by UHE protons and nuclei, sources at red-shift larger than $z>1$ can be observed only in terms of cosmogenic neutrinos \cite{Heinze:2015hhp,Berezinsky:2016feh}. Therefore a lack in the UHE neutrino flux could also be accommodated invoking a lack of sources at high red-shift.  

\section{Gamma Rays}
	
While neutrinos reach the observer without being absorbed, high energy photons and electrons/positrons colliding with astrophysical photon backgrounds (CMB and EBL) produce electromagnetic cascades (EMC) through the processes of pair production (PP, $\gamma+\gamma_{CMB,EBL}\to e^{+}+e^{-}$) and Inverse Compton Scattering (ICS, $e+\gamma_{CMB,EBL}\to \gamma + e$). While PP is characterised by a threshold the ICS process does not. From this simple observation follows that once a cascade is started by a primary photon/electron/positron it develops since the energy of photons produced by ICS are still above the PP threshold. The final output of the cascade, i.e. what is left behind when the cascade is completely developed, is a flux of low energy photons all with energies below the PP threshold. 

The two astrophysical backgrounds CMB and EBL against which the EMC develops are characterised by typical energies $\epsilon_{CMB}\simeq 10^{-3}$ eV and $\epsilon_{EBL}\simeq 1$ eV. Hence, the typical threshold energy scale for pair-production will be respectively\footnote{Numerical values quoted here should be intended as reference values being background photons distributed over energy and not monochromatic.} ${\mathcal E}_{CMB}=m_e^2/\epsilon_{CMB}=2.5\times 10^{14}$ eV and ${\mathcal E}_{EBL}=m_e^2/\epsilon_{EBL}=2.5\times 10^{11}$ eV; the radiation left behind by the cascade will be restricted to energies below ${\mathcal E}_{EBL}$. The cascade development has a universal nature independent of the spectrum of the initial photon/pair. It can be proved\footnote{For a recent detailed discussion of EMC development on CMB and EBL see \cite{Berezinsky:2016feh} and references therein.} that the spectrum of photons produced in the cascade, those left behind with energy below threshold, is always of the type: $n_{\gamma}(E_{\gamma})\propto E_{\gamma}^{-3/2}$ if $E_{\gamma} < {\mathcal E}_X$ and $n_{\gamma}(E_{\gamma})\propto E_{\gamma}^{-2}$ if ${\mathcal E}_X \le E_{\gamma} \le {\mathcal E}_{EBL}$, being ${\mathcal E}_X=(1/3) {\mathcal E}_{EBL} \epsilon_{CMB}/\epsilon_{EBL}$ the (average) minimum energy of a photon produced through the ICS mechanism by an electron with the minimum allowed energy ${\mathcal E}_{EBL}/2$ (see \cite{Berezinsky:2016feh,Aloisio:2017iyh} and references therein). The normalisation of the spectrum $n_{\gamma}(E_{\gamma})$ can be easily determined imposing energy conservation, i.e. the total energy of the cascading photons should correspond to the energy of the photon/pair that started the cascade. In the case of cascades started at high red-shift ($z>1$) some dependence on the energy and spectrum of the initiating particles arises because of the expansion of the universe. The propagation of UHECR in astrophysical backgrounds certainly produces EMC started by pairs and photons coming by pair-production and photopion production of UHECR on CMB and EBL. These cascades transform the energy lost by UHECR in low energy gamma-ray photons, with the characteristics discussed above, that in turn contribute to the diffuse gamma ray background (see \cite{Berezinsky:2016jys,Liu:2016brs,Berezinsky:2010xa} and references therein). Therefore, the observation of a diffuse extra-galactic gamma-ray background by the Fermi-LAT satellite \cite{TheFermi-LAT:2015ykq} can be used to constrain models of UHECR. The observed fast decrease in energy ($\propto E^{-2.4}$) of the diffuse background already limits models of pure proton composition, which maximise the production of secondary cosmogenic particles. Versions of the dip model with strong red-shift evolution of sources seem already ruled out by Fermi-LAT observations \cite{Berezinsky:2016jys,Liu:2016brs,Berezinsky:2010xa}. In figure \ref{fig8bis} (taken from \cite{Berezinsky:2016jys} left panel and \cite{Liu:2016brs} right panel) we plot the expected gamma ray background in the case of the dip model without cosmological evolution of sources in comparison with the experimental data of Fermi-LAT with two alternative models for the measurement of the diffuse background as discussed in \cite{Berezinsky:2016jys,Liu:2016brs,TheFermi-LAT:2015ykq}. 

\section{Conclusions}

We conclude by stating that the principal avenue through which UHECR studies should develop is toward a better experimental determination of the mass composition. Moreover, to better achieve this goal, drawing more precise conclusions about sources, a combined, multi-messenger, analysis is needed, in which also the observations of secondary gamma-rays and neutrinos should be taken into account. 

\bibliographystyle{varenna}
\bibliography{uhecr}

\end{document}